\documentclass[onecolumn,aps,prb,amsmath,amssymb, floatfix]{revtex4}
\usepackage[dvips]{graphicx}
\usepackage{bm}
\usepackage{color}
\begin{document}
\title{Switching Energy of Ferromagnetic Logic Bits}
\author{Behtash~Behin-Aein$^1$(behinb@purdue.edu),~
        Sayeef~Salahuddin$^2$,~
        and~Supriyo~Datta$^1$,~\\
        $^1$School of Electrical and Computer Engineering, Purdue University, West Lafayette, Indiana 47907 and NSF
Network for Computational Nanotechnology (NCN), Purdue University, West Lafayette, Indiana 47907\\
$^2$School of Electrical Engineering and Computer Science, UC Berkeley, Berkeley, California 94720}
\begin{abstract}
Power dissipation in switching devices is believed to be the
single most important roadblock to the continued downscaling of
electronic circuits. There is a lot of experimental effort at this
time to implement switching circuits based on magnets and it is
important to establish power requirements for such circuits and
their dependence on various parameters. This paper analyzes
switching energy which is dissipated in the switching process of
single domain Ferromagnets used as \emph{cascadable logic} bits.
We obtain generic results that can be used for comparison with
alternative technologies or guide the design of magnet based
switching circuits. Two central results are established. One is
that the switching energy drops significantly if the ramp time of
an external pulse exceeds a critical time. This drop occurs more
rapidly than what is normally expected of adiabatic switching for
a capacitor. The other result is
that under the switching scheme that allows for logic operations,
the switching energy can be described by a single equation in both fast and slow limits.
Furthermore, these generic results are used to
quantitatively examine the possible operation frequencies and
integration densities of these logic bits which show that nanomagnets can have
scaling laws similar to CMOS technology.
\end{abstract}
\maketitle
\section{Introduction}\label{S1}
It has been suggested\cite{Sayeef} that the use of collective
systems like a magnet can reduce the intrinsic switching energy
(that is dissipated throughout switching) significantly compared to that required for individual spins.
There is also a lot of experimental
effort\cite{Cowburn1,Cowburn2,Imre,Ney,Allwood1,Allwood2} at this
time to implement switching circuits based on magnets. There has
been some work \cite{Dmitri} on modeling magnetic circuits like
MQCA's in the atomic scale using quantum density matrix equation
but most of the work \cite{Csaba1,Csaba2,Csaba3,Csaba4,Porod} is
in the classical regime using the well known micromagnetic
simulators (OOMMF) based on the Landau-Lifshitz-Gilbert (LLG)
\cite{LLG1,LLG2,SD} equation. This paper too is based on the LLG
equation, but our focus is not on obtaining the energy requirement
of any specific device in a particular simulation. Rather it is to
obtain generic results that can guide the design of magnet based
switching circuits as well as
providing a basis for comparison with alternative technologies.\\
\indent The results we present are obtained by analyzing the
cascadable switching scheme illustrated in Fig.1 where the magnet
to be switched (magnet 2) is first placed along its hard axis by a
magnetic pulse (see `mid state' in Fig.1). On removing the pulse,
it falls back into one of its low energy states (up or down)
determined by the `bias' provided by magnet 1. What makes this
scheme specifically suited for logic operations is that it puts
magnet 2 into a state determined by magnet 1 (thereby transferring
information), but the energy needed to switch magnet 2 comes
largely from the external pulse \emph{and not from magnet 1}. This
is similar to conventional electronic circuits where the energy
needed to charge a capacitor comes from the power supply, although
the information comes from the previous capacitors. This feature
seems to be an essential ingredient needed to \emph{cascade logic
units}. To our knowledge, the switching scheme shown in Fig.1 was
first discussed by Bennett \cite{Bennett} and is very similar to
the schemes described in many recent publications (see e.g
Likharev et.al  \cite{Likharev}, Kummamuru et.al\cite{Lent}
 and Csaba et.al \cite{Csaba1}).\\
\begin{figure}[h]
  \centering
  \includegraphics[width=6cm, height=6cm]{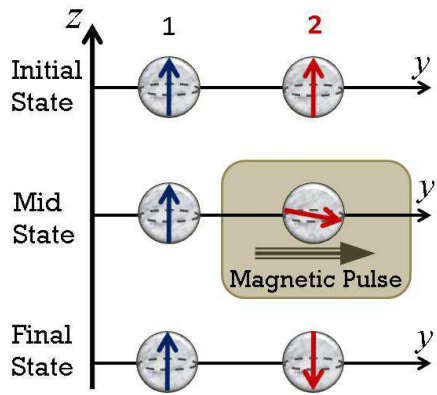}\\
  \caption{A magnetic pulse is applied to magnet 2,
  provides energy and places it along its hard axis (along $y$)
  where a small bias field due to magnet 1 can tilt it upwards or downwards
  thereby dictating its final state on removing the pulse.}
  \label{Front}
\end{figure}
\indent This paper uses the LLG equation to establish two central
results. One is that the switching energy drops significantly as
the ramp time $\tau_r$ of the magnetic pulse exceeds a critical
time $\tau_c$ given by equation \textcolor[rgb]{0.98,0.00,0.00}{(\ref{Tau_C Importance})}.
This is similar to the drop in the switching energy of an RC circuit when $\tau_r>>RC$.
But the analogy is only approximate since the switching energy for
magnets drops far more abruptly with increasing $\tau_r$. The
significance of $\tau_c$ is that it tells us how slow a pulse needs to
be in order to qualify as ``adiabatic'' and thereby reduce
dissipation significantly. Considering typical magnets used in the magnetic storage industry,
and using ramp times of a few $\tau_c$, intrinsic switching frequency of 100 MHz to 1 GHz can easily
be in the adiabatic regime of switching where dissipation is very small.\\
\indent Interestingly, we find that the switching energy for the trapezoidal
pulses investigated in this paper in both
the `fast' and `slow' limits can be described by a single equation
\textcolor[rgb]{0.98,0.00,0.00}{(Eq.\ref{General})}
which is the other central result of this paper.
\indent Later in this paper (\S\ref{S6}) we
will discuss how equations
\textcolor[rgb]{0.98,0.00,0.00}{(\ref{Tau_C Importance})}
and \textcolor[rgb]{0.98,0.00,0.00}{(\ref{General})}
can be used to guide scaling and increase switching speeds.
Furthermore these equations can be used to compare magnet based switching circuits with alternative
technologies.\\
\indent It has to be emphasized that dissipation of the external circuitry also has to be evaluated
for any new technology. A careful evaluation would require a consideration of actual circuitry to be used
(see e.g. \cite{Porod},\cite{Dmitri2})
and is beyond the scope of this paper. However following Nikonov et.al. \cite{Dmitri2}, if a wire coil is
used to produce the pulse, we can estimate the energy dissipated in creating the field $H_{pulse}$ as
$\frac{H_{pulse}^2}{2}\frac{V}{Q}$ in CGS system of units. $Q$ is the quality factor of the circuit and
$V$ is the volume over which the field extends. Depending on Q, V and $H_{pulse}$ the dissipated energy
can be much larger, comparable to or much smaller than $Ku_2V$ which sets the energy scale for the effects
considered here in this paper.\\
\indent \emph{Overview of the paper}: As mentioned before our
results are based on direct numerical simulation of the LLG
equation. However we find that in two limiting cases, it is
possible to calculate switching energy simply using the energetics
of magnetization and these limiting results are described in
sections \S \ref{S3} (dissipation with fast pulse) and \S \ref{S5}
(dissipation with adiabatic pulse) which are related to equation
\textcolor[rgb]{0.98,0.00,0.00}{(\ref{General})}. In \S \ref{S6} we use the LLG equation to show that
the switching energy drops sharply for ramp times larger than the
critical time given by equation \textcolor[rgb]{0.98,0.00,0.00}{(\ref{Tau_C Importance})}. In section
\S \ref{S7} using coupled LLG equations we analyze a chain of
inverters to show that the total dissipation increases linearly
with the number of nanomagnets thus making it reasonable to use
the one-magnet results in our paper to evaluate complex circuits,
at least approximately. Finally in section \S \ref{S8}
practical issues such as dissipation versus speed, increasing the switching speed and
scaling are qualitatively discussed in the light of these results.\\
\section{Dissipation with fast $\left(\tau_r<<\tau_c\right)$ pulse}\label{S3}
\indent Before we get into the discussion of switching energy, let
us briefly review the energetics of a magnet. The energy of a
magnet with an effective second order uniaxial
anisotropy can be described by $\frac{E}{V}=Ku_2\sin^2(\theta)$
where $\theta$ measures the deflection from the easy axis which we
take as the $z$ axis. All isotropic terms have been omitted
because they do not affect dynamics and hence dissipation of
the magnet \cite{B2}. There are two magnetic fields that control the switching
(see Fig.\ref{Front}): The external pulse $H_{pulse}$ and the bias field $H_{dc}$ due
to the neighboring magnet. Including the internal energy and the interaction
energy of magnetic moment with external fields, the energy equation reads
\begin{equation}
\label{Energy1}
 \nonumber
  \frac{E}{V} =
  -M_s\hat{m}\cdot\vec{H}_{pulse}+Ku_2\sin^2(\theta)-M_s\hat{m}\cdot\vec{H}_{dc}
\end{equation}
$M_s$ is the saturation magnetization. If the unit volume is magnetized to saturation, $M_s$ is
equivalent to the magnetic moment per unit volume. $\hat{m}$ is a unit vector in the
direction of magnetization. V is the volume of the magnet and
$Ku_2$ is the second order anisotropy constant with dimensions of
energy per unit volume. The applied field $H_{pulse}$ is along the
hard axis $\hat{y}$, the bias field $H_{dc}$ is along the easy axis $\hat{z}$ so the energy equation becomes
\begin{equation}
\label{Energy}
 \frac{E}{V} = -M_sH_{pulse}\sin(\theta)\sin(\phi)+Ku_2\sin^2(\theta)-M_sH_{dc}cos(\theta)
\end{equation}
Where $\phi$ is defined as in a standard spherical coordinate system.
Using equation \ref{Energy} we will show that dissipation with a fast pulse (small ramp time) can be written as
\begin{subequations}
    \label{DissPulseNoHdc}
    \begin{eqnarray}
   \label{Lower}
    E_{d}&=&\left(\frac{H_{pulse}}{H_c}\right)^2(2Ku_2V)  \hspace{0.5cm} \text{for} \hspace{0.5cm} H_{pulse}\le H_{c}\\
    \label{2kv}
    E_{d}&=&2Ku_2V  \hspace{0.5cm}   \text{for} \hspace{0.5cm} H_{pulse}=H_{c}\\
   \label{higher}
    E_{d}&=&\left(\frac{H_{pulse}}{H_c}\right)(2Ku_2V) \hspace{0.5cm}  \text{for} \hspace{0.5cm} H_{pulse}\ge H_{c}
    \end{eqnarray}
\end{subequations}
For reasons to be explained, under the condition of equation \ref{Lower}, logic device will not work. Nevertheless it is useful for determining dissipation in the adiabatic limit. In the equations above, $H_c\equiv\frac{2Ku_2}{M_s}$ is the minimum field necessary to put the magnet along its hard axis.
Notice that the bias field $H_{dc}$ is a dc field coming from the neighboring magnet. In practice, whether the bias field is a dc field or not, its magnitude has to be bigger than noise such that when the magnet is put along its hard axis as in Fig.\ref{Front}, the bias field can deterministically tilt the magnet towards its direction.
We will show in \S \ref{S34} that for $H_{dc}\leq 0.1 H_c$, dissipation can still be calculated using
equation \ref{DissPulseNoHdc}.\\
To derive equations \ref{DissPulseNoHdc} we find the initial and final state energies under various conditions
and evaluate the difference.
We have to emphasize that all these states essentially pertain to the energy minima (equilibrium states) i.e. they are
either the minimum of energy; or they represent a non-equilibrium
state instantaneously after the equilibrium state (minimum of energy) has changed.
Since all the fields considered here are in the $y-z$ plane and no out-of-plane field is considered,
the equilibrium states (the energy minima) will always lie in the $y-z$ plane for which $\phi = 90^\circ$.
\begin{figure}[b]
  \centering
  \includegraphics[width=8cm, height=6cm]{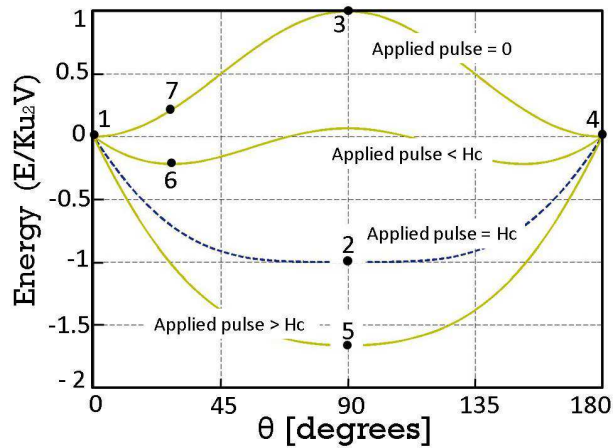}\\
  \caption{Energy landscape of the magnetization under various applied fields.
  For fast turn-on of the pulse to $H_{c}$,
  dissipation is equal to the barrier height (magnet relaxes from point 1 (or 4) to point 2).
  When the field is turned off fast, magnet relaxes from point 3 to point
  4 or 1 depending on any infinitesimal bias
  again dissipating an amount equal to the barrier height.}
  \label{Energy Landscape with no dc field}
\end{figure}
\subsection{Zero bias field ($H_{dc}=0$)} \label{S33}
Fig.\ref{Energy Landscape with no dc field} is plotted using
equation \ref{Energy} with $\phi = 90^\circ$ and $H_{dc}=0$ which
is the first case to be discussed. The different contours
correspond to different values of $H_{pulse}$.\\
\indent \emph{Derivation of equation \ref{2kv}}: Let's start with
equation \ref{2kv} which is the most important and also easiest.
Dissipation occurs both during turn-on and turn-off of the pulse
and the overall switching energy is sum of the two in general.
The dashed contour in Fig.\ref{Energy Landscape with no dc
field} corresponds to $H_{pulse}=H_c$ which is
the minimum value needed to make $\theta=90^\circ$ (point 2) the energy
minimum. For a pulse with fast ($\tau_r<<\tau_c$) \emph{turn-on},
dissipation can be calculated using equation \ref{Energy} as the
difference between the initial and the final energies which are
given by point 1 (or 4) and point 2 on the dashed contour. This
value is
\begin{equation}
\nonumber
 \label{Critical Field Dissipation On}
    E_{1\left(4\right)}-E_2=Ku_2V
\end{equation}
For a pulse with fast ($\tau_r<<\tau_c$) \emph{turn-off},
the energy contour immediately changes from the dashed one to
the uppermost one in Fig.\ref{Energy Landscape with no dc
field}. Under any infinitesimal bias, magnetization falls down the
barrier to the left (relaxing to point 1) or to the right
(relaxing to point 4) giving a dissipation of
\begin{equation}
\nonumber
 \label{Critical Field Dissipation Off}
    E_3-E_{1\left(4\right)}=Ku_2V
\end{equation}
equal to the turn-on dissipation. The switching energy
(\emph{total dissipation}) is sum of the values for turn-on and
turn-off which gives us equation \ref{2kv}.\\
\indent \emph{Derivation of equation \ref{higher}}:
This is the case with $H_{pulse} > H_c$. The bottom most energy
contour in Fig.\ref{Energy Landscape with no dc field} shows such
a situation as an example. The minimum of energy is still at
$\theta=90^\circ$ (point 5) however now the energy well is deeper.
For a pulse with fast ($\tau_r<<\tau_c$) \emph{turn-on},
dissipation is the difference between the initial and final state
energies
\begin{equation}
\nonumber
 \label{HighFieldDissipationOn}
    E_{1\left(4\right)}-E_5=(M_sH_{pulse}-Ku_2)V
\end{equation}
(Where $E_5$ is used as a generic notation for the bottom of any
well with $H_{pulse}>H_c$). For a pulse with fast
($\tau_r<<\tau_c$) \emph{turn-off}, the energy contour immediately
changes from the bottom most curve to the uppermost curve in
Fig.\ref{Energy Landscape with no dc field}. Depending on any
infinitesimal bias magnet will relax from point 3 to either point
1 or 4 dissipating the difference
\begin{equation}
\nonumber
 \label{HighFieldDissipationOff}
    E_3-E_{1\left(4\right)}=Ku_2V
\end{equation}
The switching energy is sum of the values for
turn-on and turn-off which with straightforward algebra gives us equation \ref{higher}.\\
\indent \emph{Derivation of equation \ref{Lower}}:
With $H_{pulse}<H_c$, magnetization will not align along its hard
axis ($\theta=90^\circ$). This can be seen in Fig.\ref{Energy
Landscape with no dc field} where for a pulse lower than $H_c$
there are two minima of energy not located along the hard axis.
The logic device will not work in this regime because it needs to be close to
its hard axis so that the field of another magnet can tilt it towards one
minima deterministically. Nevertheless we derive dissipation for these pulses because we
use the results in section \S \ref{S51} to show switching
energy in the adiabatic limit.
For a pulse with fast ($\tau_r<<\tau_c$) \emph{turn-on},
dissipation is the difference between the initial and final state
energies
\begin{equation}
    \label{Low Field Dissipation On}
    E_1-E_6=\left(\frac{M_sH_{pulse}}{2Ku_2}\right)^2(Ku_2V)
\end{equation}
For a pulse with fast ($\tau_r<<\tau_c$) \emph{turn-off},
the energy contour suddenly becomes the
uppermost one in Fig.\ref{Energy Landscape with no dc field}.
At that moment magnetization is still at the same $\theta$ (point
7). It follows down the barrier with the dissipation given by
\begin{equation}
    \label{Low Field Dissipation Off}
    E_7-E_1=\left(\frac{M_sH_{pulse}}{2Ku_2}\right)^2(Ku_2V)
\end{equation}
The \emph{total dissipation} is sum of the values for turn-on and
turn-off which gives us equation \ref{Lower}.
\subsection{Non-zero bias field ($H_{dc}\neq0$)} \label{S34}
In this section we show that for $H_{pulse}=H_c$, so long as
$H_{dc}\leq 0.1 H_c$ switching energy can be calculated fairly
accurately using equation \ref{2kv} considering only the effect of
$H_{pulse}$. For $H_{pulse}>H_c$ the effect of $H_{dc}$ is even
less pronounced as compared to $H_{pulse}$ and equation
\ref{higher} can be used to calculate dissipation. Again we are
interested in initial and final state energies which can be
calculated using equation \ref{Energy} with $\phi=90^\circ$.
\begin{figure}[h]
  \centering
  \includegraphics[width=8cm, height=6cm]{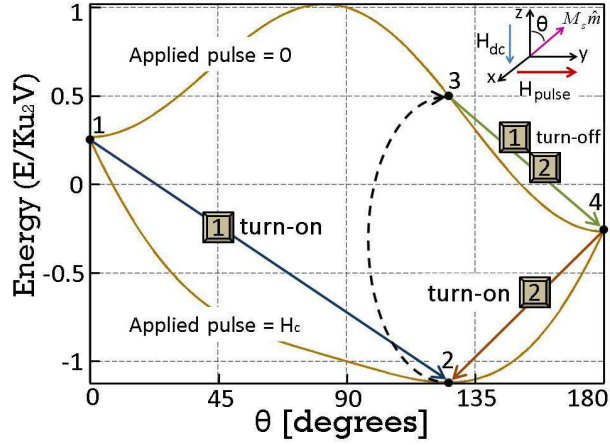}\\
  \caption{Energy landscape of magnetization with bias field $H_{dc}$ in
  the $-z$ direction for two values of the pulse: 0 and $H_{c}$.
  Upon \emph{turn-on}, if magnetization starts from $\theta=0^\circ$ (\emph{case 1}), it drops from
  point 1 ($E=+M_sVH_{dc}$) to point 2 dissipating the difference. If it starts from
  $\theta=180^\circ$, it drops from point 4 ($E=-M_sVH_{dc}$) to point 2 dissipating
  the difference. Upon \emph{turn-off}, both cases 1 and 2 drop from point 3
  to point 4 dissipating the difference.}
  \label{Energy Landscape dc}
\end{figure}
$H_{dc}$ can be positive (along $z$) or negative (along $-z$).
Fig.\ref{Energy Landscape dc} shows the energy landscape with an
$H_{dc}$ in the $-z$ direction. If $H_{dc}\neq0$ then the up and
down states (points 1 and 4) of the magnet have different initial
energies which result in two different cases to be analyzed.
\emph{Case 1} designates the situation where initial magnetization
(point 1) and $H_{dc}$ are in the \emph{opposite} direction.
\emph{Case 2} designates the situation where initial magnetization
(point 4) and $H_{dc}$ are in
the \emph{same} direction.\\
\indent For a pulse with fast ($\tau_r<<\tau_c$) \emph{turn-on},
\emph{case 1} dissipates the difference between points 1 and 2 and
\emph{case 2} dissipates the difference between points 4 and 2.
When the pulse is suddenly turned off, in both cases magnetization
finds itself at point 3, drops down to point 4 and dissipates the
difference. It is not possible to give an exact closed form
expression for the value of dissipation with non-zero bias.
Instead based on numerical calculations, we show figures that
provide useful insight to conclude that
for pulses with fast ramp time the effect of bias on switching energy is negligible.\\
\indent The energy of point 2 (and subsequently point 3) depicted
in Fig.\ref{Energy Landscape dc} changes as the relative magnitude
of $H_{dc}$ and $H_c$ are changed. We like to know how dissipation
changes as a function of the ratio $\frac{H_{dc}}{H_c}$. The
numerical results are plotted in Fig.\ref{Dbi} using equation
\ref{Energy}.
\begin{figure*}[h]
 \centering
  \includegraphics[width=18cm, height=6cm]{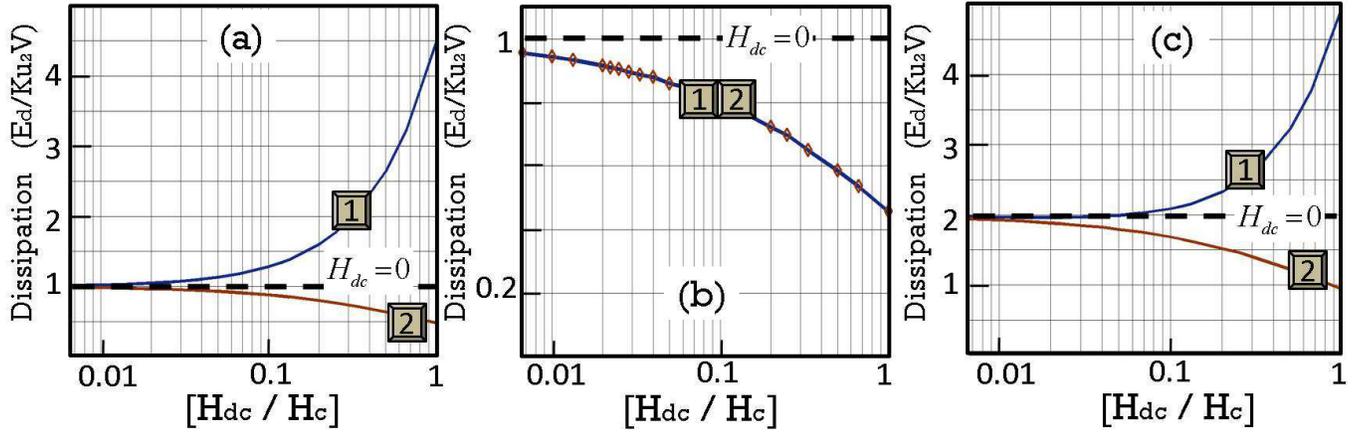}\\
  \caption{(a) Shows the \emph{turn-on} dissipation with non-zero bias.
  Cases 1 and 2 correspond to different initial directions of magnetization (see Fig.\ref{Energy Landscape
  dc}). The dashed line depicts the value of dissipation with zero bias. (b) Shows the \emph{turn-off} dissipation with non-zero bias.
  Both cases 1 and 2 dissipate the same amount (see Fig.\ref{Energy Landscape dc}). (c) Shows the total dissipation with non-zero bias.
  Notice that for relevant (small) values of $\frac{H_{dc}}{H_c}$, total
  dissipation of both cases 1 and 2 is close to the value $2Ku_2V$ which is the same as the case
  with infinitesimal bias.}
   \label{Dbi}
\end{figure*}
Fig.\ref{Dbi}a shows that for a pulse with fast \emph{turn-on} and
small values of $\frac{H_{dc}}{H_c}$, both cases dissipate about
$Ku_2V$. As this ratio is increased, the energy separation between
points 1 and 2 (see Fig.\ref{Energy Landscape dc}) increases and
that of points 4 and 2 decreases which results in higher
dissipation of \emph{case 1} and lower dissipation of \emph{case 2}.
Fig.\ref{Dbi}b shows the dissipation for a pulse with fast
\emph{turn-off} which is less than the barrier height $Ku_2V$ and
is expected because under the presence of $H_{dc}$, after turn-on,
magnetization ends up closer to the final state (see
Fig.\ref{Energy Landscape dc}) as compared to the case where
$H_{dc}=0$ (see Fig.\ref{Energy Landscape with no dc field}). The
switching energy is sum of the dissipation values for turn-on and
turn-off plotted in Fig.\ref{Dbi}c. For $H_{dc}=H_c$ the bias
field $H_{dc}$ alone can switch the magnet and it is completely an
unwanted situation \cite{ASB}. Note that for practical purposes,
values of $H_{dc}$ are small compared to $H_c$ (for instance
$H_{dc}\le 0.1 H_{c}$) and the switching energy is more or less
about $2Ku_2V$ which gives us equation \ref{2kv}. For
$H_{pulse}>H_{dc}$ the effect of bias is even less pronounced and
switching energy can be calculated using equation \ref{higher}.
\section{Dissipation with adiabatic $\left(\tau_r>>\tau_c\right)$ pulse}\label{S5}
\indent We have seen in section \S \ref{S3} that for pulses with
fast ramp times, the effect of bias ($H_{dc}$) is negligible for
$H_{dc}\le 0.1 H_{c}$ and switching energy is obtained fairly
accurately even if we set $H_{dc}=0$. By contrast for pulses with
slow ramp time, switching energy can be made arbitrarily small for
$H_{dc}=0$ and the actual switching energy is determined
entirely by the $H_{dc}$ that is used. In this section we will first show
why the switching energy can be arbitrarily small for $H_{dc}=0$ and then
show that for $H_{dc}\neq0$ it will saturate in \emph{case 1} but
can be made arbitrarily small in \emph{case 2}\cite{Disag}. Two points are in order.
First, the analysis presented here is exact in the absence of noise. If thermal noise
is present the analysis may not be true in general and needs to be modified accordingly.
Second, if in the process of switching, a bit of information is destroyed as in two inputs and one output gates (e.g. AND/OR),
then there will be a finite switching energy even for adiabatic switching.

\begin{eqnarray}
    \label{DissAdiaSummary}
    \label{Case1}
    E_{d}&=&\left(\frac{2H_{dc}}{H_{c}}\right)^p
    (2Ku_2V),\hspace{0.2cm}\left(\text{case 1: $H_{dc}$ and initial magnetization in the opposite direction}\right)
    \\
    \label{Case3}
    E_{d}&\rightarrow&0,\hspace{3cm}\left(\text{case 2: $H_{dc}$ and initial magnetization in the same direction}\right)
\end{eqnarray}
\subsection{Zero bias field ($H_{dc}=0$)}\label{S51}
\begin{figure*}[t]
 \centering
  \includegraphics[width=18cm, height=6cm]{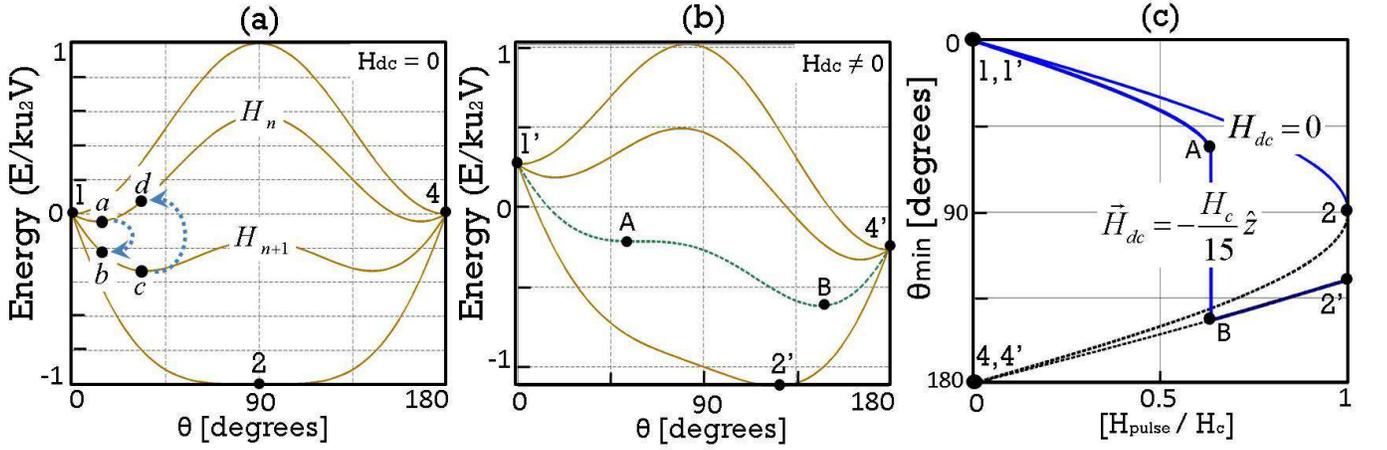}\\
  \caption{
  (a) Energy landscape of magnetization as pulse is increased from $0$ (top
  curve) to $H_c$ (bottom curve)
  with $H_{dc}=0$.
  (b) Energy landscape of magnetization as pulse is increased from 0 (top
  curve) to $H_c$ (bottom curve) with $H_{dc}\neq0$ in the $-z$ direction.
  (c) Adiabatic progression of ground state in the presence of a bias field $H_{dc}$ in the $-z$ direction.
  Figure shows those values of $\theta$ which minimize energy as the pulse is adiabatically
  ramped from 0 to 1 and back to 0. Consider the $H_{dc}\neq0$ case. If the magnetization starts at point $1^\prime$,
  it moves to point $A$ along the solid line, then suddenly drops down to point B; as $H_{pulse}$ is further increased,
  it moves to point $2^\prime$ where $H_{pulse}=H_c$. As $H_{pulse}$ is decreased back to 0, magnetization moves along the solid
  line from $2^\prime$ to $B$ and then along the dashed line to point $4^\prime$. Figures \ref{Adiabatic Progression} b and c can be used in conjunction for better understanding.}
  \label{Adiabatic Progression}
\end{figure*}

Gradual \emph{turn-on} of the pulse corresponds to increasing the
pulse in many small steps. Fig.\ref{Adiabatic Progression}a shows
the energy landscape. As the field is gradually turned-on the
energy contours change little by little from top to bottom. The minimum
of energy gradually shifts from point 1
(or 4) to point 2. Magnetization hops from one minimum of energy to
the other. But why is it that gradual turn-on of the pulse
dissipates less than sudden turn-on?\\
 \indent If the external pulse is turned on to $H_c$ in $N$ equal steps, we show that
 there is equal amount of dissipation at each step.
 Then total dissipation is $N$ times that of each step.
 We show that dissipation of each step is proportional to
 $\frac{1}{N^2}$; hence as the number of steps increases, dissipation
 decreases as $\frac{1}{N}$ and in the limit of $N\rightarrow\infty$, $E_d\rightarrow 0$ (this is not unlike
 a similar argument that has been given for charging up a capacitor adiabatically\cite{Cavin}).
 At each step when the pulse is increased by $\Delta H = \frac{H_c}{N}$, the dissipated energy is
 the difference between initial and final state energies.
Such a situation is illustrated in Fig.\ref{Adiabatic
Progression}a where $a$ denotes a minimum on an energy contour
corresponding to $H_n$ (magnitude of the pulse after $n$ steps).
When the pulse is stepped up to $H_{n+1}$, magnetization suddenly
finds itself at point $b$ (initial state) and falls down to $c$
(final state). Note that dissipation is $E_{b}-E_{c}$ and not
$E_{a}-E_{c}$. This is because when the field suddenly changes
from $H_n$ to $H_{n+1}$, magnet has not had time to relax and
dissipate energy. Here we use $E_b$ and $E_c$ as generic notations
for initial and final energy of any step. $E_{b}$ can be found by
finding the $\theta$ which corresponds to point $a$ (the minimum
of energy with $H_{pulse}=H_n$) and substituting it in equation
\ref{Energy} with $H_{pulse}=H_{n+1}$. With straightforward algebra
we get $E_b = -(M_sV)H_{n+1}\left(\frac{M_sH_n}{2Ku_2}\right)+
(Ku_2V)\left(\frac{M_sH_n}{2Ku_2}\right)^2$. Equation \ref{Low Field
Dissipation On} can be used to calculate
$E_{c}=-\left(\frac{M_sH_{n+1}}{2Ku_2}\right)^2(Ku_2V)$. Using the
identities $H_{n+1}=H_n+\Delta H$ and $\Delta H=\frac{H_c}{N}$,
the dissipated energy per step is obtained as
\begin{equation}
\nonumber
\label{DON} E_d^{step}=E_b-E_c=Ku_2V\left(\frac{1}{N^2}\right)
\end{equation}
\indent For gradual \emph{turn-off} consider points $c$,$d$ and
$a$. When $H_{pulse}=H_{n+1}$, magnetization is at $c$ and after
the pulse is decreased by one step to $H_n$, it finds itself at
$d$, falls down to $a$ dissipating the difference $E_d-E_a$.
$E_{d}$ can be found by
finding the $\theta$ which corresponds to point $c$ (the minimum
of energy with $H_{pulse}=H_{n+1}$) and substituting it in equation
\ref{Energy} with $H_{pulse}=H_n$. We get
$E_d=-(M_sV)H_n\left(\frac{M_sH_{n+1}}{2Ku_2}\right)+(Ku_2V)\left(\frac{M_sH_{n+1}}{2Ku_2}\right)^2$.
Again equation \ref{Low Field Dissipation On} can be used to give
$E_{a}=-\left(\frac{M_sH_n}{2Ku_2}\right)^2(Ku_2V)$.
Using the identities $H_{n+1}=H_n+\Delta H$ and $\Delta
H=\frac{H_c}{N}$, we obtain for the dissipated energy per step
\begin{equation}
\nonumber
\label{DOFF} E_d^{step}=E_d-E_a=Ku_2V\left(\frac{1}{N^2}\right)
\end{equation}
\indent The switching energy is sum of the dissipation values for
\emph{turn-on}: $E_d=\frac{Ku_2V}{N}$ and \emph{turn-off}:
$E_d=\frac{Ku_2V}{N}$ which in the limit of $N\rightarrow\infty$,
tends to 0 ($E_d\rightarrow 0$).
\subsection{Non-zero bias field ($H_{dc}\neq0$)}\label{S52}
\indent For \emph{turn-on} let's consider \emph{case 1} first
where initial magnetization and $H_{dc}$ are in opposite
directions (point $1^\prime$ in Fig.\ref{Adiabatic Progression}b).
As the field is gradually turned-on, magnetization starts from
point $1^\prime$ and hops from one minimum of energy to the next.
Increasing the number of steps brings the minima closer to each
other so that magnetization stays in its ground state while being
switched. However when magnetization gets to point $A$, situation
changes. At that point the energy barrier which formerly separated
the two minima on the two sides disappears. Magnetization falls
down from point A to B and dissipates the energy difference. This
sudden change in the minimum of energy occurs no matter how slow
the pulse is turned on and causes the switching energy to saturate
so long as $H_{dc}\neq0$. Quantitatively this can be seen by
plotting $\theta_{min}$ vs. $H_{pulse}$ (Fig.\ref{Adiabatic
Progression}c) using equation \ref{Energy}. When the left solid
curve is traced from $\theta_{min}=0$, it is evident that there is a
discontinuous jump in the $\theta_{min}$ values which minimize energy
when the pulse is increases from $0$ to $H_c$ in infinitesimal
steps. This discontinuity goes away only when $H_{dc}=0$ (right
solid curve). In \emph{case 2}, magnetization starts from point
$4^\prime$, i.e. $\theta_{min} = 180^\circ$ (see Fig.\ref{Adiabatic
Progression}b and c), gets to point B at which there is \emph{no}
sudden change of minimum and as the pulse is increased further to
$H_c$, it gradually moves to point $2^\prime$. During
\emph{turn-off} in both cases 1 and 2, magnetization gradually
moves
 from (see Fig.\ref{Adiabatic Progression}c)
 point $2^\prime$ to B and then finally to point $4^\prime$ all
 along staying in its minimum of energy with no discontinuity. Dissipation tends to zero as
 the pulse is turned off in infinitesimal steps.\\
\begin{figure}[t]
 \centering
  \includegraphics[width=8cm, height=6cm]{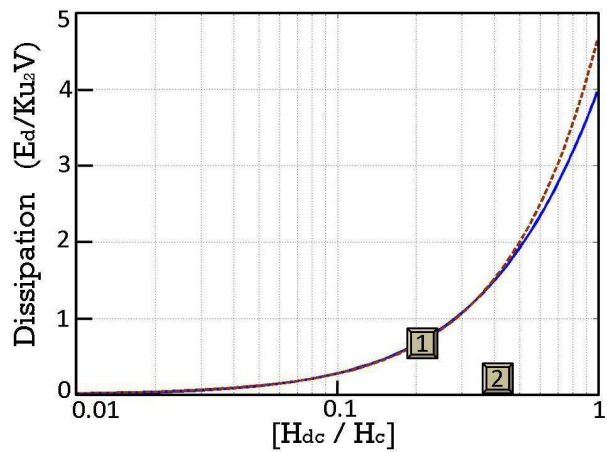}\\
  \caption{Shows the total dissipation under adiabatic switching with non-zero bias.
  There is no dissipation associated with \emph{case 2} and dissipation of \emph{case 1}
  for small (relevant) values of $\frac{H_{dc}}{H_c}$ is less than the barrier height $Ku_2V$.
  The dashed line is plotted using equation \ref{ADis}.}
  \label{Adiabatic Dissipation}
\end{figure}
\indent In the slow limit the entire dissipation is determined by
the energy difference between points $A$ and $B$, $E_A-E_B$ in
Fig.\ref{Adiabatic Progression}b. For a given $H_{dc}$, one
has to find that particular value of $H_{pulse}$ for which the
local energy maximum in the middle disappears which means that the second derivative of energy with respect
to $\theta$ must be zero (no curvature). Since magnetization has been in the minimum of energy while getting
to point $A$, first derivative of energy with respect to $\theta$ must also be equal to zero. Under these
conditions, the value of $\theta$ at $A$ and
subsequently $E_A$ can be found using equation \ref{Energy}. $E_B$ can be
found as the true minimum of energy from equation \ref{Energy} where the first derivative of energy with respect to $\theta$
is zero but the second derivative is not.
What affects $E_A-E_B$ is the relative magnitude of $H_{dc}$ and $H_c$.
It is not possible to give an analytical closed form expression
for this saturating value of dissipation. Instead we've
numerically plotted dissipation versus $\frac{H_{dc}}{H_c}$ (solid
curve in Fig.\ref{Adiabatic Dissipation}). For small values of
$\frac{H_{dc}}{H_c}$, dissipation can be written as
\begin{equation}
\label{ADis}
E_{d}=\left(\frac{2H_{dc}}{H_c}\right)^{p}(2Ku_2V),\left(p=1.23\right)
\end{equation}
Where the value of $p$ is obtained by an almost perfect fit to the
solid curve for $H_{dc}\le 0.1H_c$. The dashed curve is plotted
using equation \ref{ADis}. As is evident from Fig.\ref{Adiabatic Dissipation}, this equation is
fairly accurate. There is some digression from the actual value of dissipation for large values of
$\frac{H_{dc}}{H_c}$ which are not of practical interest especially $\frac{H_{dc}}{H_c}=1$ for which $H_{dc}$
alone can switch the magnet and is completely an unwanted situation\cite{ASB}.\\
\indent It is important to note that the switching energy in the
adiabatic limit is case dependent. For \emph{case 1}, it is given by equation \ref{ADis}
and it is not zero as it might have been expected for dissipation in the adiabatic limit.
Interestingly if $p$ was equal to 1, the dissipation would be equal to the energy
difference between initial and final states (see points $1^\prime$
and $4^\prime$ in Fig.\ref{Adiabatic Progression}b). However the
actual value is significantly smaller.\\
\indent  Dissipation in both the fast and slow limits can
be casted into a single equation
\begin{equation}
E_{d}=\left(\frac{\tilde{H}}{H_c}\right)^p(2Ku_2V) \label{General}
\end{equation}
In the fast limit, $\tilde{H}$ is the magnitude of the pulse while
in the slow limit, $\tilde{H}$ is related to the magnitude of the
small bias field as states above. $Ku_2V$ is the height of the
anisotropy energy barrier separating the two stable states of the
magnet, and has to be large enough so that the magnet retains its
state while computation is performed without thermal fluctuations
being able to flip it. The retention time for a given $Ku_2V$ can
be calculated using\cite{Street,Neel,Brown}
$t_r=t_0e^{\frac{Ku_2V}{kT}}$ where $t_0^{-1}$ is the
attempt frequency with the range $10^9-10^{12}{s}^{-1}$
\cite{Neel,Gaunt} which depends in a nontrivial fashion on
variables like anisotropy, magnetization and damping.\\
\section{Magnetization dynamics: single magnet}\label{S6}
\indent Thus far we've shown switching energy in the two limiting
cases of $\tau_r<<\tau_c$ and $\tau_r>>\tau_c$. To understand how
switching energy changes in between and also how fast it decreases
we need to start from the LLG equation which in the Gilbert form
reads:
\begin{equation}\label{LLG1}
\frac{d\vec{M}}{dt}=-|\gamma|\vec{M}\times\vec{H}
+\frac{\alpha}{|\vec{M}|}\vec{M}\times\frac{d\vec{M}}{dt}
\end{equation}
And in the standard form reads:
\begin{equation}\label{LLG2}
(1+\alpha^2)\frac{d\vec{M}}{dt}= -
|\gamma|(\vec{M}\times\vec{H})-
\frac{\alpha|\gamma|}{|\vec{M}|}\vec{M}\times(\vec{M}\times\vec{H})
\end{equation}
$\gamma$ is the gyromagnetic ratio of electron and its magnitude
is equal to $2.21\times10^5 {(rad.m)(A.s)}^{-1}$ in SI and
$1.76\times10^7 {(rad)(Oe.s)}^{-1}$ in CGS system of units.
$\alpha$ is the phenomenological dimensionless Gilbert damping
constant. $\vec{M}$ is the magnetization. Here
$\vec{H}=\vec{H}_{ani}+\vec{H}_{pulse}$ where
$\vec{H}_{ani}=\frac{2Ku_2}{M_s}m_z\hat{z}$. In general $\vec{H}$
can be derived as the overall effective field:
$\vec{H}=-\frac{1}{M_sV}\vec{\nabla}_mE$.\\
\indent The following expressions are all equivalent statements of
dissipated power \cite{SunWang,Error}:
\begin{equation}
\label{DissipationRate}
P_d=\vec{H}\cdot\frac{d\vec{M}}{dt}=\frac{\alpha}{|\gamma||\vec{M}|}\left|\frac{d\vec{M}}{dt}
\right|^2=\frac{\alpha|\gamma|}
    {(1+\alpha^2)|\vec{M}|}\left|\vec{M}\times\vec{H}\right|^2
\end{equation}
The dissipated power has to be integrated over time to give the
total dissipation. In general, LLG can be solved numerically using
the
method. To obtain generic results that are the same for
 various parameters, we recast LLG and the dissipation rate into a dimensionless
 form.  This will also show the significance of $\tau_c$ and demonstrate why for
ramp times exceeding $\tau_c=1$, there is a significant drop in dissipation.\\
\indent Using scaled variables $\vec{m}=\frac{\vec{M}}{M_sV}$ and
$\vec{h}=\frac{\vec{H}}{H_c}$ equation \ref{LLG2} in dimensionless form can be written as
\begin{figure*}[t]
  \centering
  \includegraphics[width=8cm, height=6cm]{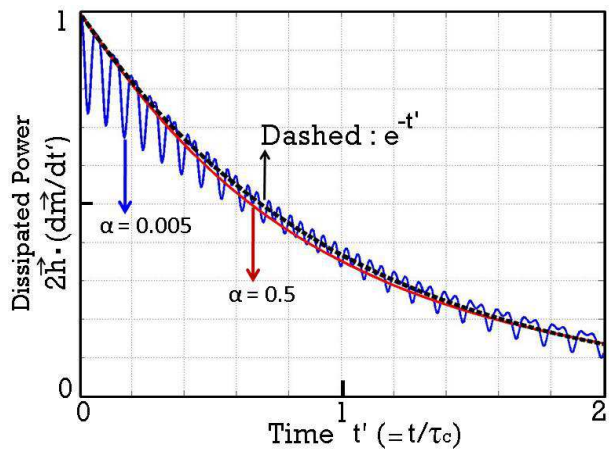}\\
  \caption{Solid lines show the dissipated power $2\vec{h}\cdot\frac{d\vec{m}}{dt^\prime}$ under
  an instantaneous turn-on of $H_{pulse}$ to $H_c$ for $\alpha=0.005$ and $\alpha=0.5$.
  Dashed line shows an exponential decay $e^{-t^\prime}$. This figure
  shows that although the value of $\alpha$ changes the time (with real dimensions) at which
 the dissipated power decreases to $1/e$ through changing $\tau_c$, it does not affect the
 functional form of the decay which is more or less an exponential decay even if $\alpha$ changes
 by 2 orders of magnitude.}
  \label{tau}
\end{figure*}
\begin{equation}\label{LLGn}
\frac{d\vec{m}}{dt^\prime}= -\frac{1}{2\alpha}(\vec{m}\times\vec{h})-\frac{1}{2}\vec{m}\times(\vec{m}\times\vec{h})
\end{equation}
where $t^\prime =\frac{t}{\tau_c}$ with $\tau_c$ given by equation \ref{Tau_C Importance}.
The energy dissipation normalized to $Ku_2V$ can be written as
\begin{equation}\label{Dissipated Energy}
\frac{E_d}{Ku_2V} = \frac{1}{Ku_2V} \int{\vec{H}\cdot\frac{d\vec{M}}{dt}dt} =
\int{2\vec{h}\cdot\frac{d\vec{m}}{dt^\prime}dt^\prime}
\end{equation}
To estimate the time constant involved in switching a magnet it is instructive to
plot the integrand
$2\vec{h}\cdot\frac{d\vec{m}}{dt^\prime}=\frac{\tau_c}{Ku_2V}\left(\vec{H}\cdot\frac{d\vec{M}}{dt}\right)$
appearing above in equation \ref{Dissipated Energy} assuming a step function for $H_{pulse}$ and obtaining
the corresponding $\frac{d\vec{m}}{dt^\prime}$ from equation \ref{LLGn}. Note that the integrands
die out exponentially for a wide range of $\alpha$'s from $0.005$ to $0.5$. In other words, all the curves (ignoring the oscillations) can be approximately described by $e^{-t^\prime}=e^{\frac{-t}{\tau_c}}$ thus suggesting that the approximate time constant is $\tau_c$
\begin{equation}
\tau_c=\frac{\left(1+\alpha^2\right)}{2\alpha (|\gamma| H_c)}
\label{Tau_C Importance}
\end{equation}
\begin{figure*}[t]
  \centering
  \includegraphics[width=18cm, height=6cm]{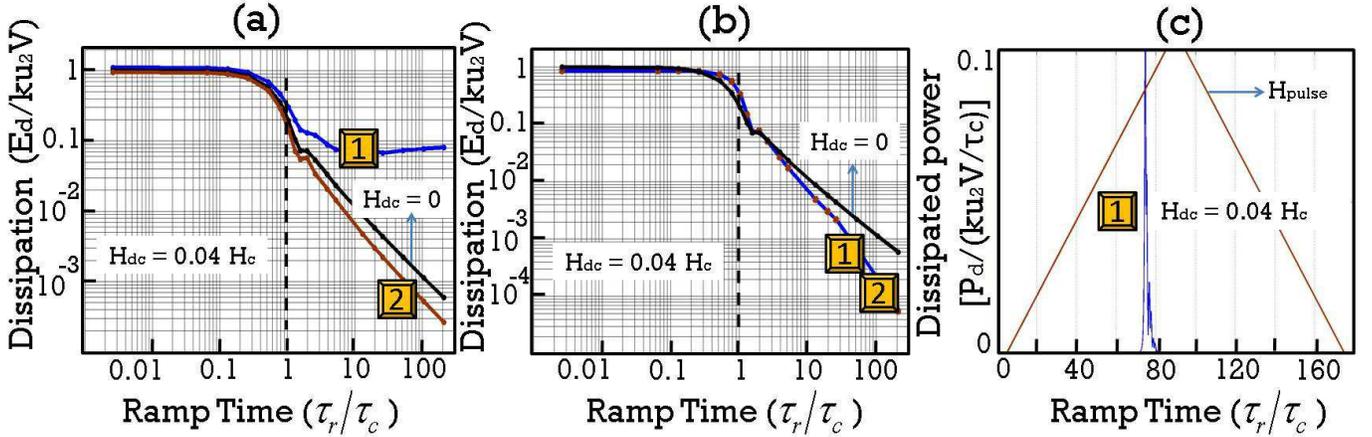}\\
  \caption{(a) \emph{Turn-on} dissipation versus ramp time. As ramp time is increased, dissipation in
  \emph{case 2} decreases arbitrary but it saturates in \emph{case 1}. In both cases
  there is a significant drop in dissipation once the ramp time exceeds $\tau_c$.
  (b) \emph{Turn-off} dissipation versus ramp time. In both cases dissipation can be made arbitrarily small by
  increasing the ramp time. Again there is a significant drop in dissipation as ramp time exceeds $\tau_c$.
  (c) Dissipated power vs. ramp time. This figure shows that in the slow limit of
  switching, for \emph{case 1} that has a saturating switching energy,
   the dissipated power essentially occurs during \emph{turn-on}. This fact
  was discussed earlier in Fig.\ref{Adiabatic Progression}b,c as the dissipation between points A and B during turn-on.
  If adiabatic limit of switching is really reached, then the dissipated power in this figure will become a very sharp spike.}
  \label{Ramp Time}
\end{figure*}
\indent This is more evident from Fig.\ref{Ramp Time} where we show the energy dissipation for pulses with
different ramp times. The dissipated energy drops when $\tau_r$ exceeds $\tau_c$ as we might expect,
but the drop is sharper than an RC circuit. Needless to say, the dissipation values
calculated from LLG equation for the two limits of fast pulse
$\left(\tau_r<<\tau_c\right)$ and adiabatic pulse
$\left(\tau_r>>\tau_c\right)$ are consistent with the values
calculated using energetics previously. Fig.\ref{Ramp Time}a shows the \emph{turn-on} dissipation
where \emph{case 1} has saturated and \emph{case 2} goes down as
ramp time is increased. The curve in the middle is the case
with infinitesimal bias $H_{dc}=0$ and it is just provided for reference.
Fig.\ref{Ramp Time}b shows the \emph{turn-off} dissipation where
both cases 1 and 2 dissipate arbitrarily small amounts as the ramp
time is increased. With slow pulses, overall switching energy of \emph{case 2}
is very small and the entire switching energy of \emph{case 1}
essentially occurs during \emph{turn-on} which is illustrated in
Fig.\ref{Ramp Time}c. This dissipation was discussed in section \S \ref{S52};
and it is associated with the sudden fall down from point A to B
(see Fig.\ref{Adiabatic Progression}b,c).
It has a saturating nature and will never become zero.
As $H_{pulse}$ is applied more and more gradually, the dissipated power
in Fig.\ref{Ramp Time}c becomes narrower and taller. In the true adiabatic
limit it will become a delta function occurring for one particular value of $H_{pulse}$.\\
\section{Magnetization dynamics: chain of inverters}\label{S7}
\begin{figure*}[h]
  \centering
  \includegraphics[width=18cm, height=6cm]{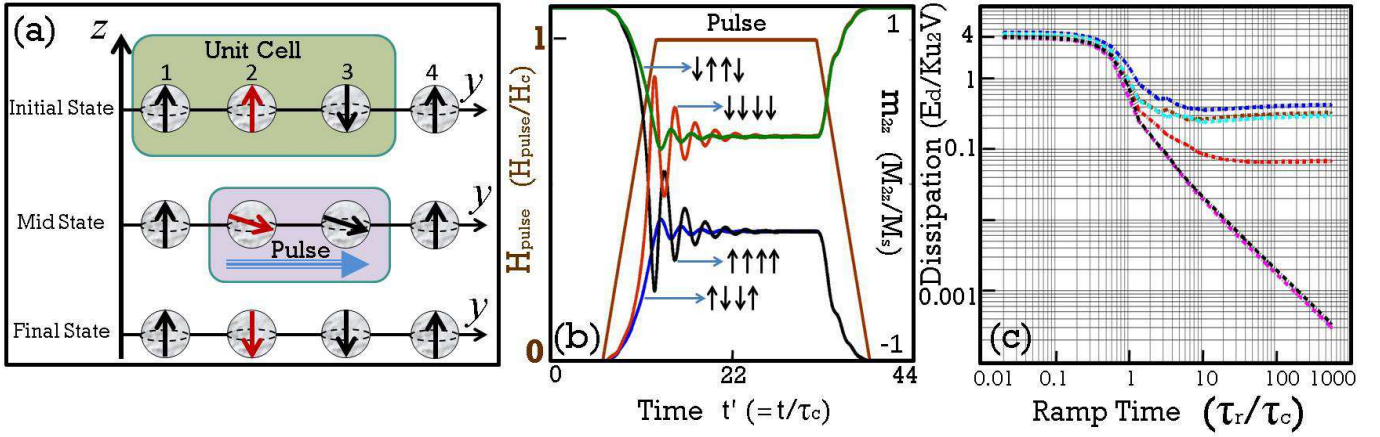}\\
 \caption{(a) An array of identical nanomagnets with uniaxial anisotropy
  and easy axis along $z$
  coupled together via dipolar coupling which can be operated as
  a 3 phase inverter chain. Initially the 4 magnet array can be randomly in any of the
  16 possible states. A unit cell is composed of 3 magnets with the real information stored in
  magnet 1 in the initial state. A $y$ pulse provides energy and puts magnets 2
  and 3 in the mid state thereby shutting off the $z$
  field of magnet 3 on 2, so that field of magnet 1 can
  deterministically tilt magnet 2 downwards. Upon removing the pulse,
  magnet 2 relaxes down in the final state.
  (b) LLG simulation of coupled system of Fig.\ref{InvChainDiss}a. This figure shows the
  proper operation of the clocking scheme by showing the normalized magnetization of magnet
  2 along its easy axis for various initial configurations.
  (c) Dissipation of the array as a function of ramp time. There are ${4\choose2}=6$
        physically distinct configurations out of 16 possible states. The dissipation is lower if the initial configuration             minimizes the energy of dipolar interaction. Assigning binary 1 to $\uparrow$ and binary 0 to $\downarrow$
        the 6 curves (from highest to lowest) represent these configurations: (1)0,15 (2)1,7,8,14 (3)3,12
        (4)2,4,11,13 (5)6,9 (6)5,10}
  \label{InvChainDiss}
\end{figure*}

Fig.\ref{InvChainDiss}a shows an array of spherical nanomagnets (MQCA) that
interact with each other via dipole-dipole
coupling\cite{Jackson}. The objective is to determine the
switching energy if we are to switch magnet 2 according to the
state of magnet 1\cite{ASB}. In section \S \ref{S71} we will show
a clocking scheme under which propagation of information can be
achieved and basically shows how magnets can be used as
\emph{cascadable logic} building blocks. In section \S \ref{S72},
we briefly go over the method and equations used to simulate the
dynamics and dissipation of the coupled magnets. In section \S
\ref{S73} we analyze the dissipation of the chain of inverters
where we show that after cascading the magnetic bits, dissipation
changes linearly with the number of magnets that the pulse is
exerted on. This shows that the switching energy of larger more complicated
circuits can be calculated using the one-magnet results presented in this paper at least
approximately.\\
\subsection{Clocking scheme}\label{S71}
\indent In the introduction we mentioned that in the clocking
scheme the role of the clock field is to provide energy whereas
field of another magnet acts as a guiding input. Using a clock we
can operate an array of exactly similar magnets as a chain of
inverters. Fig.\ref{InvChainDiss}a shows a 3 phase inverter chain
where the unit cell is composed of 3 magnets. Each magnet has two
stable states showed as \emph{up} and \emph{down} in the figure.
We want to switch magnet 2 according to the state of magnet 1.
First consider only magnets 1 and 2. We've already explained (see
section \S \ref{S1}) how magnet 1 can determine the final state of
magnet 2. But what happens if more magnets are
present?\\
\indent Consider magnets 1, 2 and 3. Just like magnet 1, magnet 3
also exerts a field on magnet 2 and if it is in the opposite
direction can cancel out the field of magnet 1. To overcome this,
we apply the pulse to magnet 3 as well thereby diminishing the
exerted $z$ field of magnet 3 on magnet 2 so that magnet 1 becomes
the sole decider of the final state of magnet 2. In the process
the data in magnet 3 has been destroyed (it will end up wherever
magnet 4 decides). It takes 3 pulses to transfer the bit (in an
inverted manner) in magnet 1 to magnet 4. Magnet 4 has been
included because it affects the dissipation of magnet 3 through
affecting its dynamics. Inclusion of more magnets to the right or
left of the array will not change the quantitative or qualitative
results of this paper. Next we'll briefly go over the method used
to simulate the chain of inverters.
\subsection{Numerical simulation of the chain of
inverters}\label{S72}
Equations \ref{LLGn} (with $\alpha=0.1$) and \ref{Dissipated Energy} are used to
simulate the dynamics and dissipation of each magnet respectively.
The overall scaled (divided by $H_c$) magnetic field $\vec{h}$ of equation \ref{LLGn}
for each magnet at each instant of time is modified to
\begin{equation}\label{ShiftRegisterH}
\vec{h}=\frac{\vec{H}_{pulse}+\vec{H}_{ani}+\vec{H}_{dip}}{H_c}
\end{equation}
composed of the applied pulse:
\begin{equation}\label{Applied pulse}
\vec{H}_{pulse} = H_{pulse}\hat{y}
\end{equation}
the anisotropy (internal) field of each magnet:
\begin{equation}\label{Anisotropy field}
\vec{H}_{ani}=\frac{2Ku_2}{M_s}m_z\hat{z}
\end{equation}
and exerted dipolar fields of other magnets which in general in CGS system of
units reads
\begin{equation}\label{Dipolar field}
\vec{H}^j_{{dip}}=\sum_{n\neq
j}\frac{3\left(\vec{\mu}_n\cdot\vec{r}_{nj}\right)\vec{r}_{nj}-\vec{\mu}_nr^2_{nj}}{r^5_{nj}}
\end{equation}
All field values are time dependent. Here $j$ denotes any one
magnet and $\mu_n$ runs over magnetic moments of the other
magnets. Though this equation can be simplified for an array of magnets
along the same line, in this form it can be used for more complicated arrangement of magnets.
Fig.\ref{InvChainDiss}b shows the LLG simulations of the
chain of inverters where magnet 2 is switched solely according to
the state of magnet 1 irrespective of its history or the state of
magnets 3 and 4.
\subsection{Dissipation of the chain of inverters with one
application of the pulse}\label{S73}
Fig.\ref{InvChainDiss}c shows dissipation of the entire array
after one application of the pulse as a function of ramp time. The
pulse is exerted on magnets 2 and 3 which accounts for the
$4Ku_2V$ value in the fast limit. This essentially points out that
after cascading these logic building blocks, dissipation changes
linearly with the number of magnets.\\
\indent In the slow limit, depending on the initial configuration,
dissipation will be affected. The 4 magnet array can initially be
in any of its 16 possible states. Some configurations saturate and
some don't. Here the field of magnet 1 plays the role of the bias
field $H_{dc}$ for magnet 2 and the field of magnet 4 is like
another bias field on magnet 3 which accounts for the 3 groups of
curves in Fig.\ref{InvChainDiss}c. The upper curves correspond to
the situation where initial magnetization of both magnets 2 and 3
are opposite to the fields exerted from magnets 1 and 4
respectively. The middle curves correspond to only one of magnets
2 or 3 initially being opposite to the exerted fields of magnet 1
or 4 respectively. The lower curves correspond to both magnets 1
and 3 initially being in the same direction as the exerted fields
from magnets 2 and 4 respectively.\\
\indent An added complication is the field of the other neighbor
(magnet 3) which is diminished in the $z$ direction but has a
non-negligible $y$ component exerted on magnet 2. All this $y$
directed field does is to wash away a tiny bit the effect of the
field of magnet 1 which has little bearing on the qualitative or
quantitative results as illustrated in Fig.\ref{InvChainDiss}c.
\section{Discussion and practical considerations}\label{S8}
\subsection{Dissipation versus speed}\label{S61}
\indent The speed of switching can be increased by increasing the
magnitude of the external pulse $H_{pulse}$ above $H_c$. Larger
fields will dissipate more energy but have the advantage of
aligning the magnet faster during the turn-on segment but are of
no use for increasing the speed of the turn-off segment because
the magnet relaxes to its stable state under its own internal
field. If $H_{c}=\frac{2Ku_2V}{M_sV}$ can be altered, then it is a
better idea to increase $H_c$ and always set $H_{pulse}=H_c$. This
way the speed of switching is increased by shortening the time of
both turn-on and turn-off segments.\\
\subsection{Increasing the switching speed}\label{S62}
\indent Consider equation (\ref{Tau_C Importance}). Increasing $\alpha$ shortens the
switching time constant (note that $\alpha$ is usually less than 1); however this parameter is not very controllable in experiments. $|\gamma|$ is a physical constant and cannot be altered. So to increase the switching speed, one has to increase $H_{c}=\frac{2Ku_2V}{M_sV}$. Thermal stability of a magnet requires $Ku_2V$ to be larger than a certain amount for the desired retention time. For instance with an attempt frequency of about 1GHz (see the discussion at the end of \S\ref{S5}) and $Ku_2V$ of about 0.5 eV, magnet is stable for about 0.5 seconds which is large enough because switching takes place in the nano-second scale. A higher retention time requires higher $Ku_2V$. Once $Ku_2V$ is set because of stability requirements, the only way to increase $H_c$ is to decrease $M_sV$. Assuming that volume is magnetized to saturation, $M_sV=N_s\mu_B$ is the magnetic moment of the magnet. $N_s$ is the number of spins giving rise to the magnetization and $\mu_B$ is Bohr magneton. So decreasing $M_sV$ translates to making the magnet smaller or decreasing its saturation magnetization.\\
\indent The discussion just presented is similar to the theory of scaling in CMOS technology where decreasing the capacitance causes an increase in the switching speed by decreasing the $RC$ time constant. With the same operating voltage, smaller capacitance results in lower number of charges stored on the capacitor. In the case of CMOS, as $C$ decreases, energy dissipated i.e. $0.5CV^2$ also decreases. In the case of magnet however, energy dissipation is fixed around $2Ku_2V$ so for a lower $M_sV$, dissipation of the Ferro-magnetic logic element (already very small) is \emph{not} altered; however one might be able to reduce the dissipated energy in the external circuitry since it needs to provide the energy for a shorter period of time.
Again we should emphasize that a thorough analysis of external dissipation
also has to be done. This has to do with generating the external source of energy for switching. In the case of MQCA circuits this is done by running currents through wires and generating magnetic fields. In principle, spin transfer torque phenomena or electrically controlled multi-Ferroicity could also be used to provide the source of energy. These methods would also have energy dissipation associated with them.
\subsection{Integration density}
\indent A complete lay-out circuit is necessary to properly evaluate the integration density of logic circuits made
of magnets. For example fringing fields and unwanted
cross talks have to be taken into account. External circuitry will take up space. Efficient methods have to be developed to porperly address these issues. One component of the lay-out is the magnetic logic bit itself which we discuss here.
The barrier height $E_b=Ku_2V$ between the stable states of a magnet
can be engineered by adjusting $Ku_2$ (anisotropy constant) and
$V$ (volume). Increasing the anisotropy constant is of great interest
for the magnetic storage industry because it allows stable magnets of smaller
volume that translates to higher densities.
Many experiments report $Ku_2$ values on the order of a few $10^7 erg/cm^3$\cite{fab1,Wu,Perumal}.
This results in stable magnets with volumes of only 10s of $nm^3$; which means that stable
magnets can be made as small as a few $nm$ in each dimension. Even though a complete lay-out is necessary, nevertheless
these numbers are very promising and could potentially result in very high integration densities.
\section{Conclusion}
In this paper we analyzed the switching energy of single domain
nanomagnets used as cascadable logic building blocks. A magnetic
pulse was used to provide the energy for switching and a bias
field was used as an input to guide the switching. The following
conclusions can be drawn from this study.\\
\indent (1) Through analyzing the complete dependence of the
switching energy on ramp time of the pulse, it was concluded that
there is a significant and sharp drop in dissipation for ramp
times that exceed a critical time given by equation \ref{Tau_C
Importance} whose significance is separating the energy dissipation characteristic of a fast pulse
(small ramp time) and energy dissipation characteristic of a slow pulse (big ramp time) .\\
\indent (2) The switching energy can be described by a single
equation (equation \ref{General}) in both fast and slow limits for trapezoidal pulses
analyzed in this paper. In
the fast limit the effect of the bias field or equivalently the
field of neighboring magnet in MQCA systems is negligible so long as the bias field is less than 10th
of the switching field of the magnet. In the slow limit however, dissipation is largely determined by
the value of the bias field.\\
\indent (3) By evaluating switching energy of both one magnet and
a chain of inverters for MQCA systems, it was shown that the
switching energy increases linearly with the number of magnets so
that the one magnet results provided in this paper can be used to
calculate the switching energy of larger more complicated
circuits, at least approximately.\\
\indent (4) Practical issues such as dissipation versus speed, increasing the switching speed and scaling were discussed qualitatively. It was concluded that by proper designing,
Ferromagnetic logic bits can have scaling laws similar to the CMOS technology.\\
\indent Noise was not directly included in the models; however we
took it into account indirectly: thermal noise is the limiting
factor on the anisotropy energy $Ku_2V$ (that determines the magnet's thermal stability) of a magnet which we
discussed thoroughly. Thermal noise also limits the lowest
possible magnitude of the bias field  (or equivalently coupling
between magnets in MQCA systems). We've provided the results for a wide range of bias values.
More thorough discussions of dissipation in the external circuitry can be found in references \cite{Porod},\cite{Dmitri2}.


\begin{thebibliography}{100}
\bibitem{B2} The isotropic terms, like the demagnetizing field for a sphere
or the Weiss field have not been taken into account because they can be expressed as $\lambda \vec{M}$
where $\vec{M}$ is the magnetization and $\lambda$ is an scalar.
Based on the LLG equation, fields of this form will \emph{not} alter the
dymanics of magnetization and have no bearing on dissipation.
\bibitem{ASB}In principle magnet A can switch magnet B
 unidirectionally with no need for an external pulse given that:
$Ku_{2,A}V_A>\frac{\left(M_{sA}V_A\right)\left(M_{sB}V_B\right)}{r^3}>Ku_{2,B}V_B$
holds. This entails designing circuits with magnets of different
parameters (e.g. volume) so no two magnets in the circuit can have
the same parameters; not to mention the complexities caused by the
fields exerted from other neighbors.
\bibitem{Disag} This essentially states that
 it is not a matter of principle that interacting magnets must have saturating
 dissipation as was previously suggested \cite{Csaba3}. This is also shown
 by analyzing the dissipation of a chain of coupled
 magnets where for some configurations dissipation arbitrarily
 goes down whereas for some other configurations it saturates. The concepts behind
 such effects can easily be traced back to the two cases (1 and 2) discussed in
 the context of one magnet and a bias field $H_{dc}$.
\bibitem{Error} Note that the the term $\vec{M}\cdot\frac{d\vec{H}_{applied}}{dt}$ should not
have been included in equation 4 of reference 1. However this will
not change the results of that paper since this term was assumed
to be zero anyways.
\end{thebibliography}
\end{document}